\documentclass[12pt,a4paper]{article}
\usepackage{axodraw}
\usepackage{graphicx}
\usepackage{amsmath}
\usepackage{amsfonts}
\usepackage{amssymb}
\newcommand{\vs}{\vspace{-0.25cm}}

\newcommand{\mev}{\,\mathrm{MeV}}
\newcommand{\fmd}{\,\mathrm{fm}^{-3}}

\newcommand{\dif}{\mathrm{d}}

\begin{document}
\hfill TUM/T39-02-01
\begin{center}

{\Large
\textbf{Chiral dynamics of nuclear matter at finite temperature}\footnote{
Work supported in part by BMBF, GSI and DFG.} 
}

\bigskip

\bigskip
S. Fritsch$\,^a$, N. Kaiser$\,^a$  and W. Weise$\,^{a,b}$\\

\bigskip

{\small
$^a$\,Physik Department, Technische Universit\"{a}t M\"{u}nchen, D-85747
Garching, Germany\\

\smallskip

$^b$\, ECT$^*$, I-38050 Villazzano (Trento), Italy\\

\smallskip

{\it email: nkaiser@physik.tu-muenchen.de}}

\end{center}

\bigskip

\begin{abstract}
We extend a recent three-loop calculation of nuclear matter in the systematic 
framework of chiral perturbation theory to finite temperatures $T$. The 
contributions from one- and two-pion exchange diagrams which cause nuclear 
binding and saturation at $T=0$ are included for $T>0$ in the density and 
temperature dependent free energy per particle, $\bar F(\rho,T)$. The so-called
anomalous $2\pi$-exchange contribution $\bar A(\rho,T)$ (with no counterpart 
in the ground state energy density at $T=0$) is consistently included. The 
calculated pressure isotherms display the familiar first-order liquid-gas phase
transition of isospin symmetric nuclear matter with a critical point at $T_c
=25.5\,$MeV and $\rho_c = 0.09\fmd$. The too high value of the critical 
temperature originates from the strong momentum dependence of the underlying 
single-particle potential $U(p,k_{f0})$ near the Fermi-surface. We also 
consider pure neutron matter at $T>0$ in the same framework and find fair 
agreement with sophisticated many-body calculations for neutron densities 
$\rho_n <0.2\,\fmd$. 
\end{abstract}

\bigskip


\bigskip

\bigskip

The analysis of data from low-energy heavy-ion collisions in the regime of
nuclear fragmentation has lead to the picture that heated nuclear matter
undergoes a first-order phase transition from a liquid-like state to a 
vaporized gas state \cite{csernai,durand}. This liquid-gas phase
transition of isospin symmetric nuclear matter is in fact very similar to that 
of the familiar van-der-Waals gas, with a corresponding critical temperature 
of $T_c \simeq  (16-18)\,$MeV \cite{durand} and a critical density of $\rho_c =
(0.06 -0.07)\fmd$ \cite{durand}. Clearly, the dynamical description of this
phase transition is an important topic in any microscopic calculation of
nuclear matter. In the $\sigma\omega$-mean field model of Serot and Walecka
\cite{walecka} nucleons are described as Dirac-quasiparticles moving in 
self-consistently generated scalar and vector mean fields and a critical 
temperature of $T_c \simeq 19\,$MeV is typically found \cite{kapusta}. As an 
other example, the sophisticated many-body calculations of the Urbana group 
\cite{urbana} using the V14 effective NN-interaction (plus an adjustable 
three-nucleon interaction) predict a critical temperature of $T_c \simeq
18\,$MeV \cite{urbana}. For earlier work on this topic using effective Skyrme 
forces, see ref.\cite{mosel}. 

In a recent work \cite{nucmat}, we have used chiral perturbation theory for a  
systematic treatment of the nuclear matter many-body problem. In this 
calculation the contributions to the energy per particle, $\bar E(k_f)$,
originate exclusively from one- and two-pion exchange between nucleons and they
are ordered in powers of the Fermi momentum $k_f$ (modulo functions of $k_f/
m_\pi$). It has been demonstrated in ref.\cite{nucmat} that the empirical 
saturation point and the nuclear matter compressibility $K\simeq 255\,$MeV can 
be well reproduced at order ${\cal O}(k_f^5)$ in the chiral expansion with just
one single momentum cut-off scale of $\Lambda \simeq 0.65\,$GeV which 
parameterizes the necessary short range dynamics. Most surprisingly, the 
prediction for the asymmetry energy, $A(k_{f0})=33.8\,$MeV, is in very good 
agreement with its empirical value. Furthermore, as a nontrivial fact pure 
neutron matter is predicted to be unbound and the corresponding equation of 
state agrees roughly with that of sophisticated many-body calculations for low 
neutron densities $\rho_n \leq 0.25\,$fm$^{-3}$. In a subsequent work 
\cite{pot}, the (complex-valued) momentum and density dependent single-particle
potential $U(p,k_f)+i\,W(p,k_f)$ (i.e. the average nuclear mean field) has been
calculated in the same framework. It was found that chiral $1\pi$- and $2\pi
$-exchange give rise to a potential depth for a nucleon at the bottom of the 
Fermi sea of $U(0,k_{f0})=-53.2\mev$. This value is in very good agreement with
the depth of the empirical optical model potential and the nuclear shell model
potential. 

Given the fact that many properties of nuclear matter can be well described
by chiral $\pi N$-dynamics treated (perturbatively) up to three loop order it 
is natural to consider in a further step finite temperatures $T$ in order to 
check whether the first-order liquid-gas phase transition of nuclear matter is 
correctly reproduced by this particular dynamics. Such an investigation is the
subject of the present paper. 

For the relatively low temperatures $T\leq 30\,$MeV which are of interest in 
this context one can safely neglect effects from thermal pions or thermally 
excited nucleon-antinucleon pairs. As a consequence, nucleons can be treated
non-relativistically and the new parameter, the temperature $T$, enters only
through the nucleons' thermal occupation probabilities given by a Fermi-Dirac
distribution.

In ref.\cite{nucmat} the diagrammatic calculation of the energy density at
$T=0$ (as a function of the particle density $\rho$) has been organized in the
number of so-called medium insertions. The latter is a technical notation for
the difference between the in-medium and vacuum nucleon propagator (see eq.(3)
in ref.\cite{nucmat}). A thermodynamically consistent extension of such an 
ordering scheme to finite temperatures is to relate it directly to the free 
energy density $\rho\,\bar F(\rho,T)$, since its natural thermodynamical 
variables are the particle density $\rho$ and the temperature $T$. In that case
the free energy density of isospin symmetric nuclear matter consists of a sum
of convolution integrals of the form, 
\begin{eqnarray} \label{Fbar}
\rho \, \bar
F(\rho,T)&=& 4\int_0^\infty \!\! \dif p_1\, p_1 \, {\cal K}_1\,d(p_1)
+\int_0^\infty \!\!\dif p_1\int_0^\infty \!\!\dif p_2\, {\cal K}_2\, d(p_1)
d(p_2)\nonumber \\ && + \int_0^\infty \!\!\dif p_1\int_0^\infty \!\!\dif
p_2\int_0^\infty \!\! \dif p_3\, {\cal K}_3 \,d(p_1) d(p_2)d(p_3) +\rho \, \bar
{\cal A}(\rho,T)\,, \end{eqnarray}
with the corresponding kernels ${\cal K}_j$ and the anomalous contribution 
$\bar {\cal A}(\rho,T)$ to be specified.\footnote{Let us briefly motivate our 
approach. The standard procedure in field theory is to calculate the 
grand canonical partition function $Z$ or equivalently the thermodynamic 
potential $\Omega(\mu,T)=-(T/V)\ln Z$ as a function of its natural variables, 
the (non-relativistic) chemical potential $\mu$ and the temperature $T$. In the
case of nuclear matter at $T=0$ the functional relationship $\Omega(\mu,0)$ is 
however double-valued for $\mu\leq 0$. Since perturbation theory usually gives 
single-valued thermodynamic functions it is more appropriate to calculate 
first the (single-valued) energy density $\rho \, \bar E(k_f)$ of nuclear 
matter as a function of the particle density $\rho$, as done e.g. in 
refs.\cite{nucmat,lutz}. Then eq.(1) defines  a thermodynamically consistent 
extension to finite temperatures $T$, with the correct $T=0$ limit. This 
calculational scheme can also be understood such that in the 
Legendre-transformation from $\Omega(\mu,T)$ to the free energy density, 
$\rho\,\bar F(\rho,T) = \Omega(\mu,T)-\mu \,\partial\Omega /\partial \mu$, the 
derivative term is taken into account only for the non-interacting (free 
nucleon gas) part and the (static) one-pion exchange contribution, but not for
the higher order contributions coming from chiral $2\pi$-exchange.} The
quantity  
\begin{equation} 
d(p_j) = {p_j\over 2\pi^2} \bigg[ 1+\exp{p_j^2 -2M \tilde \mu
\over 2M T} \bigg]^{-1} \,,
\end{equation}
denotes the density of nucleon states in momentum space. It is the product of 
the temperature dependent Fermi-Dirac distribution and a kinematical prefactor 
$p_j/ 2\pi^2$ which has been included in $d(p_j)$  for convenience. 
$M=939\,$MeV stands for the (free) nucleon mass. The particle density $\rho$ is
calculated from the density of states in momentum space as
\begin{equation}
\rho= 4\int_0^\infty  \!\! \dif p_1\, p_1 \,d(p_1) = -\sqrt{2} \Big( {MT \over
\pi} \Big)^{3/2} \, {\rm Li}_{3/2}(-e^{\tilde \mu /T})\,, \end{equation} 
and this relationship determines the dependence of the effective one-body 
"chemical potential" $\tilde \mu(\rho,T)$ on the thermodynamical variables 
$\rho$ and $T$. The "true" chemical potential is different and given by the
formula $\mu = \bar F + \rho \,\,\partial \bar F/\partial \rho$. Concerning
eqs.(1,2,3) we are following here (partially) the approximation scheme of 
ref.\cite{lejeune}. Our approach is (by construction) thermodynamically 
consistent, since all thermodynamic quantities are derived (via standard 
relations) from the free energy density $\rho \, \bar F(\rho,T)$. The infinite
series Li$_\nu(x) = \sum_{k=1}^\infty k^{-\nu} x^k$ defines the polylogarithmic
function of index $\nu$ for $|x|<1$. 

The one-body kernel ${\cal K}_1$ in eq.(1) represents the  contribution of the
non-interacting nucleon gas to the free energy density and it reads 
\begin{equation} \label{K1}
{\cal K}_1 = \tilde \mu- {p_1^2\over 3M}- {p_1^4\over 8M^3} \,. \end{equation}
While the first two terms are standard \cite{lejeune}, the correction term, 
$-p_1^4/8M^3$, has been constructed according to the following criteria. First,
it ensures the correct $T=0$ limit for the energy per particle $\bar E(k_f)$
up to order $k_f^4$, in which $\tilde \mu(\rho,0)= k_f^2/2M $ and $\rho= 2k_f^3
/3\pi^2$. Secondly, the so-constructed kernel ${\cal K}_1$ combined with the 
non-relativistic Fermi-Dirac distribution (see eq.(2)) gives a very accurate 
approximation of the free energy density of a fully relativistic free nucleon 
gas \cite{kapusta} for the densities and temperatures of interest here. The 
factor $4$ in eqs.(1,3) counts the spin-isospin multiplicity of a nucleon.    

\bigskip

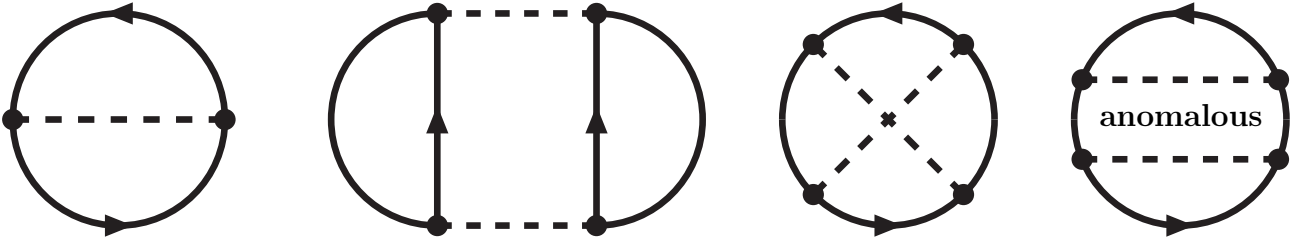
\begin{figure} 
\begin{center}
\SetWidth{2.5}

\begin{picture}(440,100)

\ArrowArc(20,50)(40,0,180)
\ArrowArc(20,50)(40,180,360)
\DashLine(-20,50)(60,50){6}
\Vertex(-20,50){4}
\Vertex(60,50){4}

\CArc(140,50)(40,90,270)
\ArrowLine(140,10)(140,90)
\DashLine(140,90)(200,90){6}
\Vertex(140,90){4}
\Vertex(200,90){4}
\CArc(200,50)(40,-90,90)
\ArrowLine(200,10)(200,90)
\DashLine(140,10)(200,10){6}
\Vertex(140,10){4}
\Vertex(200,10){4}

\ArrowArc(310,50)(40,0,180)
\ArrowArc(310,50)(40,180,360)
\DashLine(338.3,78.3)(281.7,21.7){6}
\DashLine(338.3,21.7)(281.7,78.3){6}
\Vertex(338.3,78.3){4}
\Vertex(281.7,21.7){4}
\Vertex(338.3,21.7){4}
\Vertex(281.7,78.3){4}

\ArrowArc(420,50)(40,0,180)
\ArrowArc(420,50)(40,180,360)
\DashLine(457,65)(383,65){6}
\DashLine(383,35)(457,35){6}
\Vertex(457,65){4}
\Vertex(457,35){4}
\Vertex(383,65){4}
\Vertex(383,35){4}
\Text(422,52)[]{\bf anomalous}

\end{picture}
\end{center}
\caption{The one-pion exchange Fock-diagram, the iterated one-pion exchange 
Hartree- and Fock-diagrams and the anomalous Fock-diagram. The combinatoric 
factors of these diagrams are 1/2, 1/4, 1/4 and 1/2, in the order shown. 
\label{fig1}}
\end{figure}

The (non-anomalous) contributions to the free energy density $\rho\,\bar 
F(\rho,T)$ arising from pion-exchange interactions are encoded in the kernels
${\cal K}_{2,3}$ in eq.(1). The closed vacuum diagrams related to one-pion
exchange (Fock-diagram) and iterated one-pion exchange (Hartree- and
Fock-diagrams) are shown in Fig.\,\ref{fig1}. The $1\pi$-exchange Fock-diagram
including the relativistic $1/M^2$-correction leads to the following
contribution to the two-body kernel ${\cal K}_2$:
\begin{eqnarray}{\cal K}_2^{(1\pi)} &=& {3g_A^2 \over 16 f_\pi^2} \bigg\{ 8p_1
p_2-2m_\pi^2 \ln{m_\pi^2+(p_1+p_2)^2 \over m_\pi^2+(p_1-p_2)^2} + {1\over M^2} 
\bigg[ -4p_1p_2(p^2_1+p^2_2) \nonumber \\ && + m_\pi^2(p^2_1+p^2_2)
\ln{m_\pi^2+(p_1+p_2)^2 \over  m_\pi^2+(p_1-p_2)^2 }-{2m_\pi^2 p_1p_2 (p^2_1 -
p^2_2)^2 \over [m_\pi^2+(p_1+p_2)^2][m_\pi^2+(p_1-p_2)^2]} \bigg] \bigg\} \,.
\end{eqnarray}
As in ref.\cite{nucmat}, we choose the value $g_A=1.3$ for the nucleon axial 
vector coupling constant, $f_\pi = 92.4\,$MeV denotes the weak pion decay
constant and $m_\pi=135\,$MeV stands for the (neutral) pion mass. The iterated 
$1\pi$-exchange Hartree-graph (second diagram in Fig.\,\ref{fig1}) contributes 
to the two-body kernel ${\cal K}_2$ with the term:
\begin{eqnarray} \label{K2H}
{\cal K}_2^{(H)} &=& {3g_A^4M m_\pi^2 \over 8\pi f_\pi^4} 
\bigg\{ (p_1+p_2) \arctan{p_1+p_2\over m_\pi}  \nonumber \\ && + (p_2-p_1) 
\arctan{p_1-p_2\over m_\pi}  - {5\over 8} m_\pi \ln{m_\pi^2+(p_1+p_2)^2 \over 
m_\pi^2+(p_1-p_2)^2} \bigg\} \,. \end{eqnarray} 
In this expression we have omitted the contribution of a linear divergence
proportional to the momentum cut-off $\Lambda$ (see ref.\,\cite{nucmat}). All 
such powerlike terms in $\Lambda$ are collected in eq.\eqref{Klam}. Similarly,
the iterated $1\pi$-exchange Fock-graph (third diagram in Fig.\,\ref{fig1}) 
gives rise to a two-body kernel ${\cal K}_2$ of the form:
\begin{eqnarray}{\cal K}_2^{(F)} &=& {3g_A^4M m_\pi \over 32\pi f_\pi^4} 
\bigg\{ 2p_1p_2 +m_\pi^2 \int_{|p_1-p_2|/2m_\pi}^{(p_1+p_2)/2m_\pi} \!\!{\dif 
x \over 1+2 x^2} \nonumber \\ && \times \Big[ (1+8x^2+8x^4) \arctan x-(1+4x^2) 
\arctan 2x \Big] \bigg\} \,.
\end{eqnarray} 
The additional diagrams of irreducible $2\pi$-exchange (not shown here, but 
see Fig.\,4 in ref.\cite{nucmat}) generate a contribution to the two-body
kernel ${\cal K}_2$ via the expression
\begin{equation} \label{K2ir}
{\cal K}_2^{(2\pi)} = {m_\pi^4 \over 128\pi^2 f_\pi^4} \bigg\{
I\Big({p_1+p_2 \over 2m_\pi} \Big) -I\Big({|p_1-p_2| \over 2m_\pi} \Big)
\bigg\} \,,
\end{equation}
with the function
\begin{eqnarray} I(x)&=& 3(11g_A^4-2g_A^2-1) \ln^2(x+ \sqrt{1+x^2}) \nonumber 
\\ && + 2(g_A^2-1) \Big[ g_A^2(31+22x^2) +5+2x^2\Big]\, x \sqrt{1+x^2}\ln(x+ 
\sqrt{1+x^2}) \nonumber \\ && +(7-2g_A^2+91g_A^4) x^2 +(3+14g_A^2-g_A^4) x^4  
\nonumber \\ && +\Big[ 12(15g_A^4-6g_A^2-1)x^2
+4(11g_A^4 -10 g_A^2 -1) x^4 \Big] \ln{m_\pi \over 2\Lambda} \,,
\end{eqnarray} 
obtained from solving the pion-loop integrals. The complete expression for the 
power divergences specific to cut-off regularization reads: 
\begin{equation} \label{Klam}
{\cal K}_2^{(\Lambda)} ={3\Lambda \, p_1p_2 \over 32\pi^2 
f_\pi^4 } \Big[ -10g_A^4 M +(3g_A^2+1)(g_A^2-1) \Lambda\Big] \,.
\end{equation}
The term linear in the cut-off $\Lambda$ stems from iterated $1\pi$-exchange
with a contribution of the Hartree- and Fock diagram in the ratio $4:1$. The
term quadratic in the cut-off, on the other hand, originates from irreducible
$2\pi$-exchange. Note that the kernel ${\cal K}_2^{(\Lambda)}$ in 
eq.\eqref{Klam} leads to a temperature independent contribution to free energy
per particle $\bar F(\rho,T) \sim \rho$. Therefore it is fully equivalent to 
a momentum independent NN-contact interaction. 

Next, we come to the three-body kernel ${\cal K}_3$ which incorporates the
temperature and density dependent Pauli-blocking effects in intermediate 
NN-states. The iterated $1\pi$-exchange Hartree-graph contributes to the
three-body kernel ${\cal K}_3$ in the form: 
\begin{equation} {\cal K}_3^{(H)} ={3 g_A^4M  \over 4 f_\pi^4 } \int_{|p_1-
p_2|}^{p_1+p_2} \!\! \dif q \,{q^4\over(m_\pi^2 +q^2)^2}\ln {|p_1^2-p_2^2+q^2+
2p_3 q| \over |p_1^2-p_2^2+q^2-2p_3 q|} \,,  \end{equation}
and from the iterated $1\pi$-exchange Fock-graph one finds,
\begin{eqnarray} {\cal K}_3^{(F)} &=&{3 g_A^4M  \over 16 f_\pi^4   } \bigg\{
{1\over 8 p_3^3} \bigg[ 4p_1p_3+(p_3^2-p_1^2-m_\pi^2 ) \ln{m_\pi^2+(p_1+p_3)^2
\over m_\pi^2+(p_1-p_3)^2} \bigg] \nonumber \\ && \times \bigg[ 4p_2p_3+(p_3^2
-p_2^2-m_\pi^2 ) \ln{m_\pi^2+(p_2+p_3)^2 \over m_\pi^2+(p_2-p_3)^2} \bigg] 
\nonumber \\ && + \int_{|p_2-p_3|}^{p_2+p_3}\!\!\dif q \,{ q^2 \over m_\pi^2 
+q^2} \bigg[ \ln{ |p_1+h| \over |p_1-h|} +{m_\pi^2 \over R} \ln {
|p_1 R +(p_1^2-p_3^2 -m_\pi^2)h| \over |p_1 R +(p_3^2 +m_\pi^2-p_1^2)h| } 
\bigg] \bigg\} \,,\end{eqnarray}
with the abbreviations:
\begin{equation} R =\sqrt{(m_\pi^2+p_1^2-p_3^2)^2 +4m_\pi^2(p_3^2-h^2)} \,,
\quad \quad h = {1\over 2q} (p_2^2-p_3^2-q^2) \,.
\end{equation}
Note that all integrands in representations of ${\cal K}_{2,3}$ are odd 
functions of their respective integration variable and therefore one could even
drop the absolute magnitude on the lower integration limits. We also remind
that (non-anomalous) terms involving the product of four Fermi-Dirac 
distributions are effectively absent in second order perturbation theory 
\cite{thouless}. Because of the antisymmetry of the accompanying energy
denominator under the exchange of two pairs of momenta these terms integrate to
zero.  

Next, we come to the so-called anomalous contribution $\bar {\cal A}(\rho,T)$ 
in eq.(1). This is a special feature at finite temperatures 
\cite{thouless,kohn} with no counterpart in the calculation of the ground state
energy density $\rho \, \bar E(k_f)$ at $T=0$. From the "anomalous"
Fock-diagram in Fig.\,1 we derive the following contribution to the free energy
per particle of isospin symmetric nuclear matter:  
\begin{eqnarray} \bar{\cal A}(\rho,T) &=& - { [\Omega_{1\pi}'(\rho,T)]^2 \over
2\rho\, \Omega_0'' (\rho,T)}\nonumber \\ && + {9 g_A^4 \over 8 f_\pi^4 T \rho} 
\int_0^\infty \!\!\dif p_1\int_0^\infty \!\!\dif p_2\int_0^\infty \!\! \dif p_3
\,\, d(p_1)d(p_2)[2\pi^2 d(p_2)-p_2] d(p_3)\nonumber \\ && \times  \bigg[ p_1-
{m_\pi^2\over 4p_2} \ln{m_\pi^2+(p_1+p_2)^2 \over m_\pi^2+(p_1-p_2)^2}\bigg] 
\bigg[ p_3-{m_\pi^2\over 4p_2}\ln{m_\pi^2+(p_3+p_2)^2\over m_\pi^2+(p_3-p_2)^2}
\bigg]  \,, \end{eqnarray}
with the $\tilde \mu$-derivative of the thermodynamical potential due to static
$1\pi$-exchange, 
\begin{equation} \Omega_{1\pi}'(\rho,T) =  {3 g_A^2 M \over 2f_\pi^2} 
\int_0^\infty \!\!\dif p_1\int_0^\infty \!\!\dif p_2\,\, d(p_1){d(p_2) \over
p_2} \bigg[ {(p_1+p_2)^3 \over m_\pi^2+(p_1+p_2)^2}+{(p_1-p_2)^3 \over m_\pi^2
+(p_1-p_2)^2} \bigg]\,, \end{equation}
and the second $\tilde \mu$-derivative of the free nucleon gas part, 
\begin{equation}\Omega_0'' (\rho,T)=- 4M\int_0^\infty  \!\! \dif p_1\,
\,{d(p_1) \over p_1} = \sqrt{2T}\, \Big( {M\over\pi} \Big)^{3/2} \, 
{\rm Li}_{1/2}(-e^{\tilde \mu /T})\,. \end{equation}
The first term in eq.(14) originates from taking into account the (static)
$1\pi$-exchange contribution in the Legendre-transformation from the
thermodynamical potential to the free energy density and from the perturbative
shift of the chemical potential $\tilde \mu \to  \tilde \mu-\Omega_{1\pi}'(
\rho,T)/\Omega_0'' (\rho,T)$ (for details on that procedure, see 
ref.\cite{kohn}).  We have explicitly checked that the anomalous contribution 
$\bar{\cal A}(\rho,T)$ vanishes identically at $T=0$ for all densities $\rho$. 
This vanishing, $\bar{\cal A}(\rho,0)=0$, is an automatic consequence of the
Kohn-Luttinger-Ward theorem \cite{kohn} which of course holds in our case since
the Fermi-surface and the pion-induced interactions are invariant under spatial
rotations. Furthermore, it is interesting to observe that the temperature and
density dependent anomalous contribution $\bar{\cal A}(\rho,T)$ vanishes
identically in the chiral limit $m_\pi= 0$. Note also that the anomalous
contribution $\bar{\cal A}(\rho,T)$ is (formally) of the same order in small 
momenta as the contributions to $\bar F(\rho,T)$ coming from iterated
$1\pi$-exchange. It must therefore not be neglected in a consistent and
complete calculation.     

Via general thermodynamical relations \cite{kapusta} one finally derives from 
the free energy per particle $\bar F(\rho,T)$ the pressure  $P(\rho,T)$ and the
entropy per  particle $\bar S(\rho,T)$ as:  
\begin{equation} P(\rho,T) = \rho^2 \, {\partial \bar F(\rho,T) \over \partial
\rho} \,,  \qquad \bar S(\rho,T) = -{\partial \bar F(\rho,T) \over \partial T} 
\,. 
\end{equation}
We are now in the position to present numerical results for isospin symmetric
nuclear matter at finite temperatures. We use consistently the same parameters 
as in our previous work \cite{nucmat}. There, our only parameter, the cut-off 
scale $\Lambda=646.3\,$MeV\,$\simeq7f_\pi$, has been fine-tuned to the binding 
energy per particle, $-\bar E(k_{f0}) = 15.26\,$MeV. Fig.\,2 shows the free
energy per particle $\bar F(\rho,T)$ as a function of the nucleon density 
$\rho$ for various temperatures $T= 0,10,18,25.5,30$\,MeV. The uppermost line 
is the attractive branch of the nuclear matter saturation curve at $T=0$. The
singular behavior of the free energy per particle $\bar F(\rho,T)$ for $\rho\to
0$ at $T>0$ is a generic feature (see e.g. the figures and tables corresponding
to the results of the Urbana group in ref.\cite{urbana}). The internal energy
per particle $\bar F(\rho,T)+T\,\bar S(\rho,T)$, on the other hand, approaches 
approximately the value $3T/2$ for $\rho\to 0$. 


\begin{figure}
\begin{center}
\includegraphics[scale=0.6]{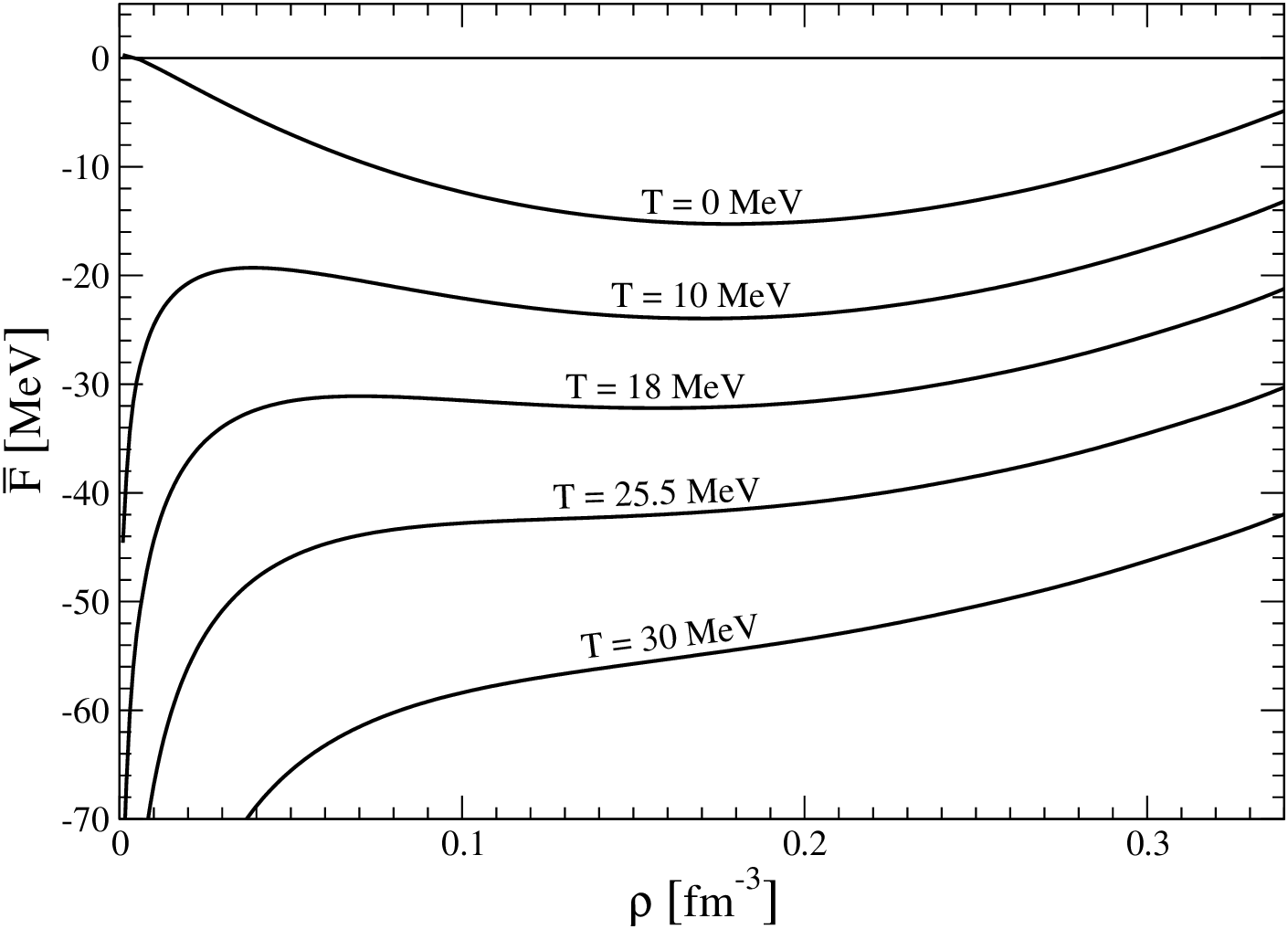}
\end{center}
\vspace{-0.2cm}
\caption{The free energy per particle of isospin symmetric nuclear matter 
$\bar F(\rho,T)$ versus the nucleon density $\rho$. Each curve is labeled with
its corresponding constant temperature $T$.}
\bigskip
\end{figure}


Fig.\,3 shows the calculated pressure isotherms $P(\rho,T)$ of isospin
symmetric nuclear matter. As it should be these curves display a first-order
liquid-gas phase transition similar to that of the van-der-Waals gas. The
coexistence region between the liquid and the gas phase (which has to be
determined by the Maxwell construction \cite{kapusta}) terminates at the
critical temperature $T_c$.  From there on the pressure isotherms $P(\rho,T)$
grow monotonically with the particle density $\rho$. In the present calculation
we find a critical temperature of $T_c= 25.5\,$MeV and a critical density
of $\rho_c = 0.09\,\fmd \simeq 0.5 \rho_0$ (with $\rho_0 = 0.178\,\fmd$, 
the predicted nuclear matter saturation density). Together with the critical 
pressure $P(\rho_c,T_c) = 0.69\,$MeV$\fmd$ this prediction of the
critical point deviates considerably from most other calculations. The too high
value of the critical temperature $T_c = 25.5\,$MeV finds its explanation 
in the strong momentum dependence of the underlying single-particle potential 
$U(p,k_{f0})$ near the Fermi-surface $p=k_{f0}$ (see Fig.\,3 in
ref.\cite{pot}). The nominal value of the effective nucleon mass at the  
Fermi-surface is $M^*(k_{f0}) \simeq 3 M$. The latter quantity determines 
via the reciprocally related density of states at the Fermi-surface completely
the low-temperature behavior of a Fermi-liquid. Since in our calculation the 
density of thermally excitable quasi-particles is too low nuclear matter has 
to be heated up to higher temperatures until it evaporates completely. More
elaborate calculations of nuclear matter in effective (chiral) field theory are
necessary in order to cure this particular problem of the large effective 
nucleon mass $M^*(k_{f0})$. Note that despite its large isospin factor 18 the
anomalous contribution $\bar {\cal A}(\rho,T)$ does practically not influence
the behavior of nuclear matter at low temperatures $T<30\,$MeV.  

As a side remark, we consider the chiral limit $m_\pi=0$. With a reduced 
cut-off scale of $\Lambda = 555.8\,$MeV (and fixed $g_A$,\,$f_\pi$,\,$M$) the 
same maximum binding energy per particle ($15.26\,$MeV) is obtained at a 
saturation density of $\rho_0 =0.145\,\fmd$.  Interestingly, the critical 
temperature $T_c\simeq 27\,$MeV remains nearly unchanged when taking the chiral
limit. This confirms the expectation that the critical temperature
$T_c$ is primarily determined by the binding energy per particle $-\bar
E(k_{f0})$ and the effective nucleon mass at the Fermi-surface $M^*(k_{f0})$.


\begin{figure}
\begin{center}
\includegraphics[scale=0.6]{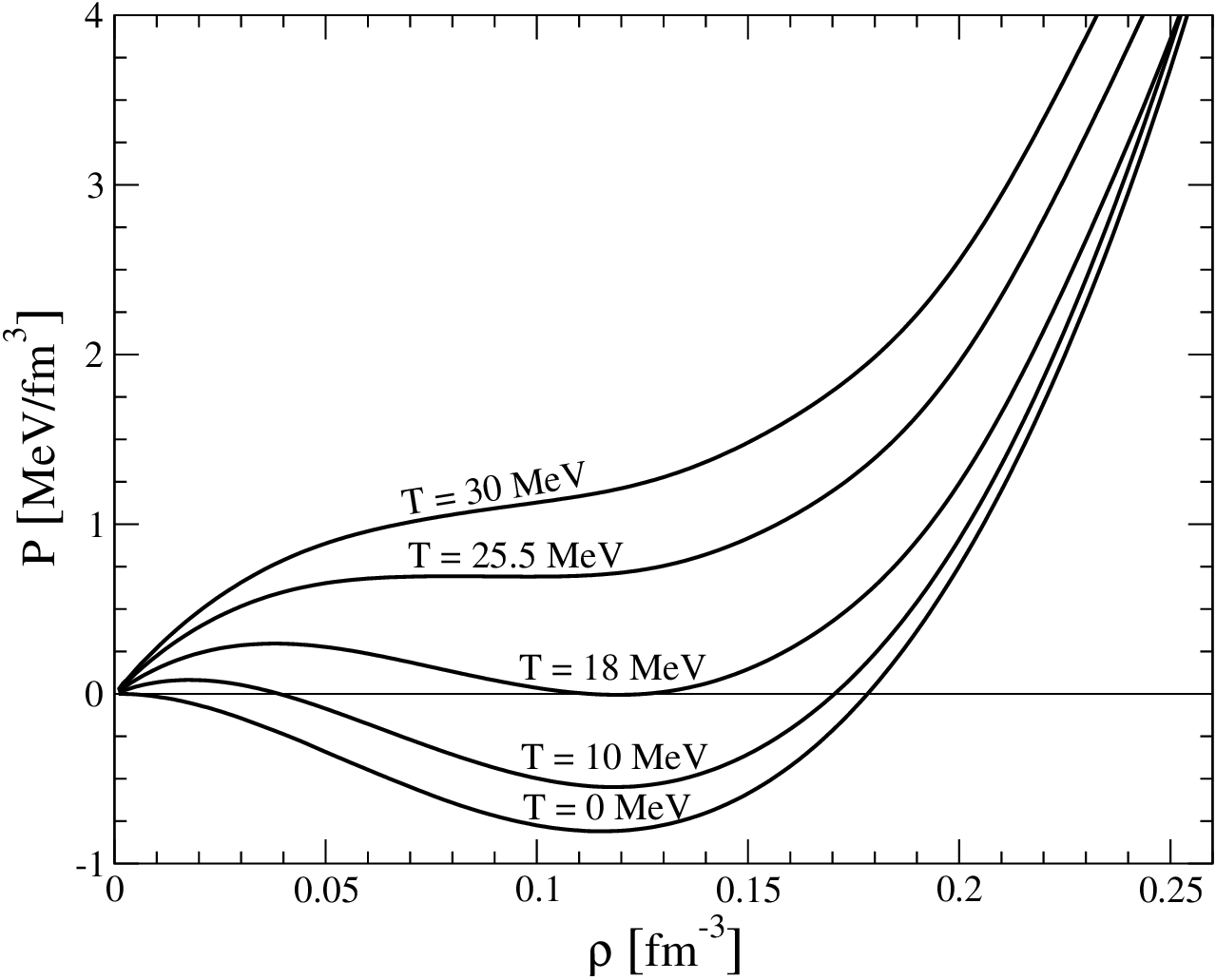}
\end{center}
\vspace{-0.2cm}
\caption{Pressure isotherms $P(\rho,T)$ of isospin symmetric nuclear matter. 
The coexistence region of the liquid and gas phase ends at the critical point 
$\rho_c \simeq 0.09 \,\fmd$ and $T_c \simeq 25.5\,$MeV.} 
\medskip
\end{figure}


The present calculation is readily adapted to the case of pure neutron matter
at finite temperatures $T$. Only the isospin factors of the pion-exchange 
diagrams (see Fig.\,1) change when switching from isospin symmetric nuclear
matter to pure neutron matter. The free energy per neutron $\bar F_n(\rho_n,T)$
takes a form analogous to eq.(1), namely: 
\begin{eqnarray} \rho_n \, \bar F_n(\rho_n,T)&=& 2\int_0^\infty \!\! \dif p_1
\,p_1\,{\cal K}_1\,d(p_1) +\int_0^\infty \!\! \dif p_1\int_0^\infty \!\! \dif 
p_2\, {\cal K}_{n,2}\, d(p_1) d(p_2)\nonumber \\ && + \int_0^\infty \!\!\dif
p_1\int_0^\infty \!\! \dif p_2\int_0^\infty \!\!\dif p_3\,{\cal K}_{n,3}\,
d(p_1)d(p_2) d(p_3) +  \rho_n\, \bar{\cal A}_n(\rho_n,T)\,.  \end{eqnarray} 
The neutron density $\rho_n= \rho /2$ is reduced by a factor 2 in comparison
to $\rho$ calculated via eq.(3) and the same applies to the one-body term 
proportional to ${\cal K}_1$ in eq.(18). The contributions from $1\pi$-exchange
and iterated $1\pi$-exchange to the neutron kernels ${\cal K}_{n,j}$ are
determined by certain relative isospin factors as,  
\begin{equation}  {\cal K}_{n,2}^{(1\pi)}=
{1\over 6} {\cal K}_2^{(1\pi)}\,, \qquad {\cal K}_{n,j}^{(H)}= {1\over 12} 
{\cal K}_j^{(H)}\,, \qquad {\cal K}_{n,j}^{(F)}= -{1\over 6} {\cal
K}_j^{(F)}\,, \quad (j=2,3)\,,  
\end{equation}
while irreducible $2\pi$-exchange leads to the expression
\begin{equation}{\cal K}_{n,2}^{(2\pi)} = {m_\pi^4 \over 768\pi^2 f_\pi^4} 
\bigg\{I_n\Big({p_1+p_2 \over 2m_\pi} \Big) -I_n\Big({|p_1-p_2| \over 2m_\pi} 
\Big)\bigg\} \,, 
\end{equation}
with the modified function 
\begin{eqnarray} I_n(x)&=&3(19g_A^4-2g_A^2-1) \ln^2(x+ \sqrt{1+x^2}) \nonumber 
\\ && - 2\Big[ g_A^4(26x^2-7) +2g_A^2(13+10x^2) +5+2x^2\Big]\, x \sqrt{1+x^2}
\ln(x+\sqrt{1+x^2}) \nonumber \\ &&  +(19+118g_A^2-257g_A^4) x^2
+(3+14g_A^2-9g_A^4) x^4  \nonumber \\ && +\Big[ 12(1+6g_A^2-15g_A^4)x^2
-4(13g_A^4 +10 g_A^2 +1) x^4 \Big] \ln{m_\pi \over 2\Lambda} \,.
\end{eqnarray} 
The expression for the power divergences specific to cut-off 
regularization changes accordingly in the case of pure neutron matter,
\begin{equation} {\cal K}_{n,2}^{(\Lambda)} =-{\Lambda \, p_1p_2 \over 64\pi^2 
f_\pi^4 } \Big[ 2g_A^4 M +(3g_A^2+1)(g_A^2-1) \Lambda\Big] \,. \end{equation}
Furthermore, for pure neutron matter the anomalous contribution gets reduced by
a relative isospin factor $1/18$, or written in an explicit formula: $\bar{\cal
A}_n(\rho_n,T)=\bar {\cal A}(2\rho_n,T)/9$ with $\bar {\cal A}(\rho,T)$ given 
in eq.(14). With this small weight factor the anomalous $2\pi$-exchange Fock
diagram becomes irrelevant for pure neutron matter.

In Fig.\,4, we show by the full lines the calculated free energy per neutron
$\bar F_n(\rho_n,T) $ for temperatures $T=10,20\,$MeV. The dashed lines in 
Fig.\,4 correspond to the many-body calculation of the Urbana group
\cite{urbana}. These curves should be considered as a representative of the
host of existing neutron matter calculations which scatter around them. In
order to demonstrate the effects of the $nn$-interaction, we show by the dotted
lines in Fig.\,4 the result of a free neutron gas (as given by the one-body
kernel ${\cal K}_1$ in eq.(18)). One observes a fair agreement of our 
calculation with the results of ref.\cite{urbana} for low neutron densities, 
$\rho_n \leq 0.2 \,\fmd$. One may also conclude from Fig.\,4 that the 
parameterfree interaction effects generated by chiral $1\pi$- and 
$2\pi$-exchange are fairly realistic in this low density region. 

In summary, we have extended our recent three-loop calculation of nuclear
matter in chiral perturbation theory to finite temperatures $T$. The 
contributions from one- and two-pion exchange diagrams are included in the
density and temperature dependent free energy per particle $\bar F(\rho,T)$. 
This guarantees thermodynamical consistency and the correct $T=0$ limit to the
energy per particle $\bar E(k_f)$. We reproduce the familiar liquid-gas phase 
transition of isospin symmetric nuclear matter. The predicted critical point, 
$T_c = 25.5 \,$MeV, $\rho_c = 0.09\,\fmd$ and $P(\rho_c,T_c) = 
0.69\,$MeV$\fmd$ lies however too high in temperature. The obvious reason for
that is the strong momentum dependence of the single-particle potential 
$U(p,k_{f0})$ \cite{pot} near the Fermi-surface $p=k_{f0}$ implying a huge 
effective nucleon mass $M^*(k_{f0})\simeq 3M$. More elaborate calculations of
nuclear matter in effective (chiral) field theory are necessary in order to 
cure this problem. Furthermore, we have considered pure neutron matter at
finite temperatures $T$ in the same framework. We have found fair agreement
with sophisticated many-body calculations for the low neutron densities,
$\rho_n \leq 0.2\,\fmd$, relevant for conventional nuclear physics.

\begin{figure}
\begin{center}
\includegraphics[scale=0.6]{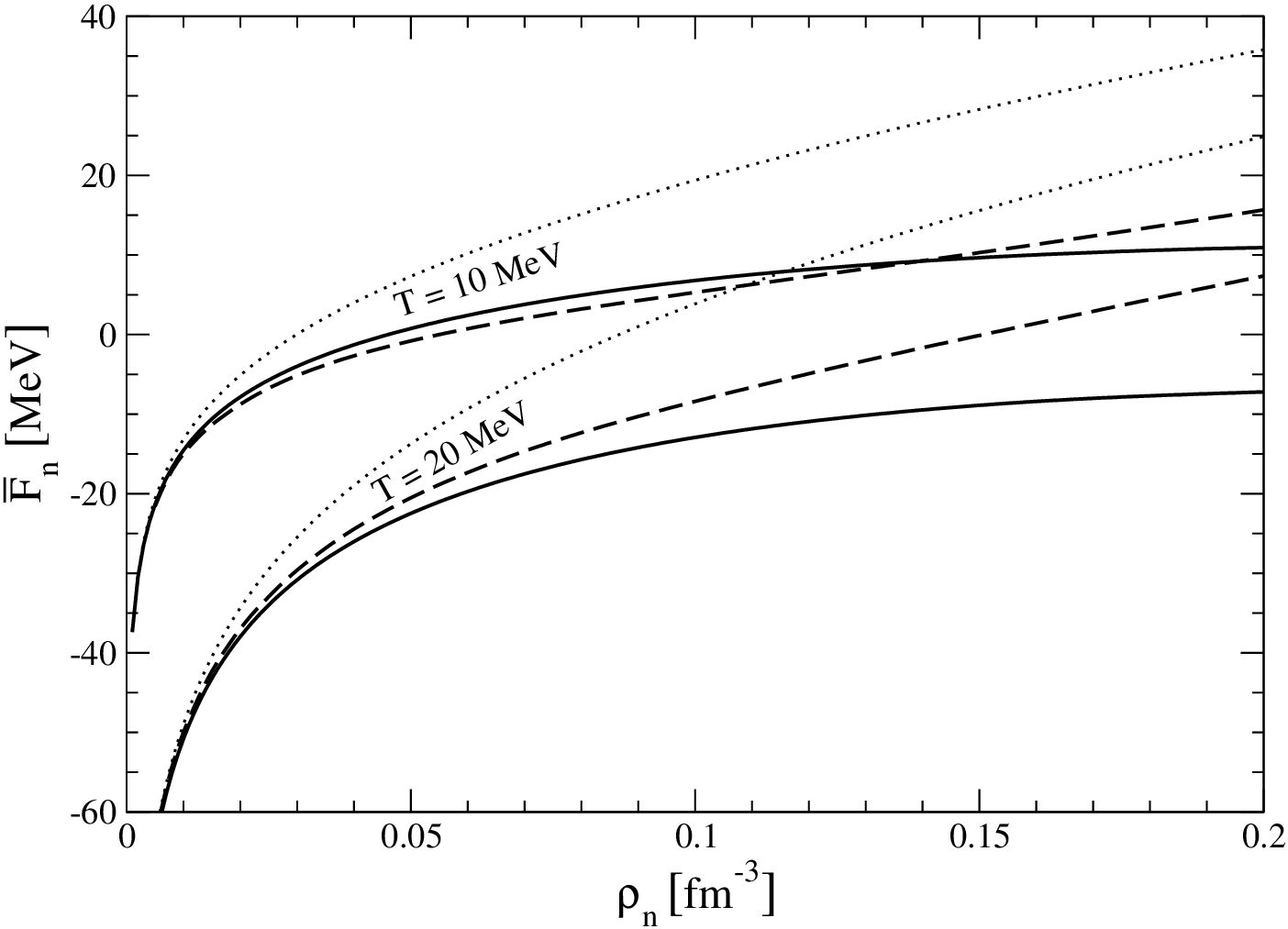}
\end{center}\vspace{-0.2cm}
\caption{Free energy per particle of pure neutron matter $\bar F_n(\rho_n,T)$. 
The full lines give the result of chiral one- and two-pion exchange. The dashed
lines correspond to the many-body calculation of ref.\cite{urbana} and the
dotted lines show the result for a non-interacting neutron gas.}
\end{figure}

\end{document}